\documentclass[
preprintnumbers,
preprint,
a4paper,
showpacs, 
amsmath, 
amssymb, 
floatfix,
nofootinbib,
]{revtex4}

\usepackage[dvipdfmx]{graphicx}

\begin{document}

\title{Uniqueness of static photon surfaces: Perturbative approach}

\author{Hirotaka Yoshino}


\affiliation{Department of Mathematics and Physics,
Graduate School of Science, Osaka City University,
Osaka 558-8585, Japan}

\preprint{OCU-PHYS-453, AP-GR-133}

\date{Submitted: July 28, 2016; Published: February 28, 2017}

%
%
\begin{abstract}
  A photon surface $S$ is defined as a three-dimensional
  timelike hypersurface such that
  any null geodesic initially tangent to $S$ continues
  to be included in $S$, like $r=3M$ of the Schwarzschild spacetime.
  Using analytic solutions to static perturbations of a Schwarzschild
  spacetime, 
  we examine whether a nonspherical spacetime
  can possess a distorted static photon surface.
  It is shown that if the region outside of $r=3M$
  is vacuum, no distorted photon surface can be present.
  Therefore, we establish the perturbative uniqueness 
  for an asymptotically flat vacuum spacetime with a static photon surface.
  It is also pointed out that
  if matter is present in the outside region, there is a possibility
  that a distorted photon surface could form.
\end{abstract}

\pacs{04.20.Cv, 04.20.Ex, 02.40.Hw, 04.25.Nx}

\maketitle

%
%

\section{Introduction}

A remarkable property of a black hole is the existence
of an event horizon. But in predicting observational image 
around a black hole, the location of
closed circular orbits of null geodesics (say, photon rings)
is even more important. 
In a Schwarzschild spacetime, a photon ring is located at $r=3M$, and
all photon rings form a
{\it photon sphere} due to spherical symmetry. More generally,  
a photon sphere  
is defined as a static timelike surface $S$
with spherically symmetric geometry such that
any null geodesic initially tangent to $S$ will remain tangent to $S$
\cite{Virbhadra:1999,Claudel:2000}.
The photon sphere has important effects
on particles/waves traveling around a static black hole,
and as a consequence, on the gravitational lensing effects \cite{Virbhadra:1999,Bozza:2002,Sahu:2012,Sahu:2013}
and black hole shadows \cite{Synge:1966} 
(see also Ref.~\cite{Grenzebach:2014}).

A related notion is the {\it photon surface},
which was proposed as generalization
of the photon sphere \cite{Claudel:2000}.
The photon surface is 
defined as a timelike hypersurface $S$ such that
any null geodesic initially tangent to $S$ continues to be
included in $S$.
The photon surface is a broader notion compared to the photon sphere:
it may be dynamical or may not be spherically symmetric.
But spacetimes possessing photon surfaces are fairly 
restricted. 
For example, a Kerr spacetime
does not possess a photon surface.
Although there are null geodesics each of which remains on a $r={\rm const.}$ 
surface in the Boyer-Lindquist coordinates,
the $r$ value depends on 
the angular momentum of null geodesics \cite{Teo:2003} (see also Sec. 5.8
of Ref.~\cite{Perlick:2004}). 
As a result, there is a photon region
in which null geodesics staying on $r={\rm const.}$ surfaces
distribute like layers.
On the inner/outer boundary of this region,
prograde/retrograde 
null geodesics rotate on the equatorial plane.
In the nonrotation limit, this photon region becomes infinitely thin 
and reduces to a photon surface. In this manner, a photon surface
is formed as a special limit of general cases.

In this paper, we consider a static photon surface in a static
vacuum spacetime. A static photon surface
naturally arises when the spacetime is spherically symmetric
with strong gravitational source.
Our question is whether a nonspherical
spacetime can possess a static photon surface or not.
Specifically, we consider a vacuum spacetime with 
a (possibly distorted) static photon surface as an inner boundary.
Here, we do not take care of
the inside region of the photon surface: There may be a nonspherical
star composed of unusual matter or  
an event horizon surrounded by 
a naked singularity,
and so on, which causes 
distortion of a photon surface/spacetime from spherical symmetry.

A partial answer was given as the uniqueness theorem 
for a {\it (generalized) photon sphere}.
The author of Ref.~\cite{Cederbaum:2014} redefined the notion of
a photon sphere as a
static photon surface on which the time lapse function
$N=\sqrt{-g_{tt}}$ is constant. 
Then, analyzing the Einstein equations in the outside region
with the boundary conditions on the photon sphere, the spacetime is
shown to be spherically symmetric. Namely, 
an asymptotically flat vacuum spacetime
that allows the presence of a photon sphere
is only the Schwarzschild spacetime \cite{Cederbaum:2014,Cederbaum:2015a}. 
This uniqueness theorem was generalized to
electrovacuum spacetimes \cite{Yazadjiev:2015a,Cederbaum:2015b}
and to other spacetimes \cite{Yazadjiev:2015b,Yazadjiev:2015c,Rogatko:2016}.

The constancy of the time lapse function $N$ on a photon sphere
would be imposed by technical reason.
The question here is what happens if we relax this condition, i.e.,
if we consider a surface
on which the time lapse function $N$ is nonconstant.
As a first step toward this direction, we adopt the
perturbative approach.
Namely, we consider a static perturbation of a Schwarzschild
spacetime and study whether a static photon surface
exists or not in a distorted Schwarzschild spacetime.

Analyses of static perturbations are useful, because if absence
of a regular solution is proved, no regular branch of static solutions
starting from the background solution exists, and therefore, a restriction
on the solution space is obtained.
For example, 
the perturbative uniqueness was shown for a large class of higher-dimensional
static black hole solutions \cite{Kodama:2004}.
Conversely, if there is a regular solution to 
a static perturbation, we notice the possibility that 
an unknown solution branch exists.
One example is the static perturbation
at the threshold of the Gregory-Laflamme instability
of black string spacetimes~\cite{Gregory:1993},
which motivated the numerical constructions
of nonuniform black string solutions in the nonlinear regime~\cite{Wiseman:2002,Kleihaus:2006}.

This paper is organized as follows.
In the next section, we summarize the geometrical
properties of the photon surface, which will be used
to judge its existence in a perturbed
Schwarzschild spacetime.
In Sec.~III, we discuss what the absence/presence
of the first-order solution can tell us on the
nonexistence/existence of a fully nonlinear solution.
In Sec.~IV, analytic solutions to
static perturbations of a Schwarzschild
spacetime are derived.
In Sec.~V, we examine whether a photon surface
exists or not in a distorted Schwarzschild spacetime.
It is shown that a photon surface vanishes
once the spacetime is distorted, if the region outside of $r=3M$ is vacuum.
We also point out that it may be possible to realize a distorted
photon surface if matter is present outside of $r=3M$.
Section VI is devoted to a conclusion.
In the Appendix, the detailed calculations on the photon surface
condition in a perturbed Schwarzschild spacetime are explained.
To simplify the notation, we use the geometrical units 
$c = G = 1$.

%
%
\section{Photon surface}

In this section, we review the geometrical properties
of a photon surface.
Let $S$ be a timelike three-dimensional hypersurface
in a spacetime $(M,g_{ab})$ and 
$n^a$ be the unit normal to $S$.
The induced metric $h_{ab}$ and the extrinsic curvature $\chi_{ab}$
of $S$ are introduced by
\begin{equation}
h_{ab}=g_{ab}-n_an_b,
\end{equation}
\begin{equation}
\chi_{ab} = h_{a}^{~c}\nabla_cn_b = \frac12\pounds_nh_{ab},
\end{equation}
where $\pounds_n$ denotes the Lie derivative with respect to $n^a$.
In Ref.~\cite{Claudel:2000}, three expressions
of the necessary and sufficient condition 
for $S$ to be a photon surface are presented.
The first expression is that affine-parametrized null geodesics of the
submanifold $S$ are simultaneously affine-parametrized null
geodesics of the spacetime $M$.
The second expression is that for arbitrary null vectors $k^a$
tangent to $S$,
\begin{equation}
  \chi_{ab}k^ak^b = 0
  \label{light-surface-condition-2}
\end{equation}
holds. The third expression is that the hypersurface $S$ is umbilical, i.e.,
\begin{equation}
  \chi_{ab} \propto h_{ab}.
  \label{light-surface-condition}
\end{equation}
Each of these three expressions is equivalent to the
condition for $S$ to be a photon surface, and
the proofs are given in Ref.~\cite{Claudel:2000}.

Here, we discuss intuitive interpretation for these conditions.
It is useful to recall the variational principle for  
affine-parametrized null geodesics. The action is given by
\begin{equation}
\mathcal{S}=\int \mathcal{L} d\lambda, \qquad \mathcal{L}=\frac12g_{ab}k^ak^b.
\end{equation}
Consider a null geodesic of $M$ whose path is included 
in $S$. If we slightly deform its path,
the action is unchanged to first order, and this holds
also when the deformation is restricted on $S$.
Therefore, the null geodesic under consideration is
also a null geodesic of the submanifold $S$.
Conversely, for a hypersurface $S$ to be a photon surface,
each null geodesic of the submanifold $S$ has to be a null geodesic of $M$
(i.e., the first expression).
This means that the action $\mathcal{S}$ must be stationary when 
its path is deformed toward outside of $S$,
i.e., in the direction of $n^a$. Since change
in the Lagrangian $\mathcal{L}$ in this deformation
is proportional to $\chi_{ab}k^ak^b$, the condition~\eqref{light-surface-condition-2} must hold (i.e., the second expression).
If the condition~\eqref{light-surface-condition} is satisfied,
the condition~\eqref{light-surface-condition-2} is obviously 
satisfied as well.
Because of the arbitrariness of $k^a$, no other form of $\chi_{ab}$
is allowed (i.e., the third expression).

Among the three expressions, the third one, Eq.~\eqref{light-surface-condition},
is used in this paper in order 
to identify the (non)existence of a photon surface in a distorted
Schwarzschild spacetime. 
In what follows, the condition~\eqref{light-surface-condition}
is referred as the
``photon surface condition''.

%
%
\section{What a perturbation can tell us}

In this paper, we study the first-order metric perturbation
of a Schwarzschild spacetime. Before starting the analysis,
we discuss what absence/presence
of a solution of a first-order perturbation can tell us
on the nonexistence/existence of a fully nonlinear solution
in this section.

In the perturbation theory, the metric
is expanded as
\begin{equation}
  g_{ab}=g^{(0)}_{ab}+\epsilon g^{(1)}_{ab}+\epsilon^2 g^{(2)}_{ab}
  +\cdots,
  \label{Eq:metric-expand-general}
\end{equation}
with a small expansion parameter $\epsilon$
and $g^{(0)}_{ab}$ is a background spacetime metric.
In our case, $g^{(0)}_{ab}$ is the Schwarzschild solution of the vacuum
Einstein equation,
\begin{equation}
  R_{ab}=0,
  \label{Eq:vacuum-Einstein}
\end{equation}
and the perturbed metric is also required to satisfy
this equation. 
Substituting
the formula~\eqref{Eq:metric-expand-general} into Eq.~\eqref{Eq:vacuum-Einstein}
and collecting the same-order terms with respect to $\epsilon$, 
the zeroth-order equation is trivially satisfied
and the first-order and higher-order equations have the following form:
\begin{subequations}
\begin{eqnarray}
  \mathcal{L}\left(g^{(1)}_{ab}\right) &=& 0;\\
  \label{Eq:1st-order}
  \mathcal{L}\left(g^{(n)}_{ab}\right) &=& \mathcal{S}_n\left(g^{(1)}_{ab},...,g^{(n-1)}_{ab}\right),~~(n=2,3,...),
  \label{Eq:nth-order}
\end{eqnarray}
\end{subequations}
where $\mathcal{L}$ is the linear operator and
$\mathcal{S}_n\left(g^{(1)}_{ab},...,g^{(n-1)}_{ab}\right)$ is sum of
the products of
$g^{(1)}_{ab},...,g^{(n-1)}_{ab}$ with $n$th order
(see \cite{Gleiser:1995} for a similar expression in
the second-order case).

In our problem, we prepare a regular 
solution of $g^{(1)}_{ab}$ 
and look for a photon surface that satisfies the photon
surface condition, $h_{ab}\propto \chi_{ab}$.
The photon surface is assumed to deviate
from $r=3M$ as $\epsilon$ is increased.
Suppose it is shown that the photon surface condition
is not compatible with any 
nontrivial first-order perturbation. Namely, we consider the
case that the only first-order solution consistent with
the presence of a photon surface is $g^{(1)}_{ab}=0$
and the photon surface does not distort up to $O(\epsilon)$.
In this situation, we can show
that the $n$'th-order perturbation also vanishes
for arbitrary $n$ using the mathematical induction.
Assuming $g_{ab}^{(i)}=0$ for all $i=1,...,n-1$, 
the source term $\mathcal{S}_n$ of Eq.~\eqref{Eq:nth-order} becomes zero and 
the solution space of $g^{(n)}_{ab}$ is same as that of $g^{(1)}_{ab}$.
Setting $\epsilon^\prime = \epsilon^{n}$, the perturbation
is effectively same as the first-order perturbation with respect to
$\epsilon^\prime$. Then, by the same reason as the first-order
case, we have $g_{ab}^{(n)}=0$ and the photon surface does not distort
up to $O(\epsilon^n)$.
Therefore, the absence of a nontrivial
first-order solution
implies the absence of a nontrivial solution at any order.
There is no regular sequence of fully nonlinear solutions
branching from the background spacetime.

Note that although any regular solution sequence branching
from the Schwarzschild solution can be ruled
out in the above manner, there remains a possibility that 
a sequence of regular spacetimes 
approaching the Schwarzschild limit in a singular way
might exist at least logically 
(see Sec.~3.3 of \cite{Kodama:2004}
for such an example in the context of the ordinary static
black hole uniqueness).
Therefore, showing 
the nonexistence of perturbative solutions
does not give a complete proof of the uniqueness
near the Schwarzschild limit in the solution space.  
Such a possibility is left open in this paper.

On the other hand, suppose a first-order solution with a distorted
photon surface can be constructed.
In this case, the existence of the first-order solution does not
guarantee the existence of a fully nonlinear solution,
because there is a possibility that the second-order (or higher-order)
perturbation is incompatible with the photon surface condition.
In other words, the existence of a first-order solution
is a necessary
condition for the existence of a fully nonlinear solution,
but not a sufficient condition. 
Therefore, we have to interpret
the existence of the first-order solution as an indication
for the possible existence
of a fully nonlinear solution.

\section{Static perturbation of Schwarzschild spacetime}

Now, we study static first-order perturbations
of a Schwarzschild spacetime. 
The background metric is given in the coordinates
$(t, r, \theta, \phi)$ as follows:
\begin{equation}
d\hat{s}^2=-e^{2\nu^{(0)}}dt^2+e^{2\mu^{(0)}}dr^2+r^2
\left(d\theta^2+\sin^2\theta d\phi^2\right),
\label{SAD1}
\end{equation}
\begin{equation}
e^{2\nu^{(0)}}=e^{-2\mu^{(0)}}= 1 - \frac{2M}{r}. 
\label{SAD2}
\end{equation}
Static distortion of a Schwarzschild spacetime
is represented by
time-independent even-parity perturbations \cite{Regge:1957}.
After a suitable gauge transformation (i.e., the Regge-Wheeler
gauge \cite{Regge:1957}), 
the metric of a distorted Schwarzschild spacetime is
written in diagonal form:\footnote{
  Although the perturbed metric in the Regge-Wheeler gauge
  has nonzero $(tr)$ component in general, this vanishes in the static case
  due to the {\it T} symmetry of the Einstein's equation (i.e., the symmetry
  under the time-reversal transformation, $t\to -t$). }
\begin{equation}
d\hat{s}^2=-e^{2\nu}dt^2+e^{2\mu}dr^2+e^{2\psi}r^2
\left(d\theta^2+\sin^2\theta d\phi^2\right), 
\label{metric_deformed}
\end{equation}
\begin{subequations}
\begin{eqnarray}
\nu &=& \nu^{(0)}+\epsilon \nu^{(1)}+\cdots,
\label{expand_nu}\\
\mu &=& \mu^{(0)}+\epsilon \mu^{(1)}+\cdots,
\label{expand_mu}\\
\psi &=& \epsilon \psi^{(1)}+\cdots,
\label{expand_psi}
\end{eqnarray}
\end{subequations}
with a small expansion parameter $\epsilon$.
The first-order functions
are expanded with the angular eigenmodes,
\begin{subequations}
\begin{align}
&\nu^{(1)}=-\mu^{(1)}=-\sum_{\ell, m}H_{\ell m}^{(1)}(r) Y_{\ell m}(\theta,\phi), 
\label{function_nu1_mu1}\\
&\psi^{(1)}= \sum_{\ell,m}K_{\ell m}^{(1)}(r) Y_{\ell m}(\theta,\phi),
\label{function_psi1}
\end{align}
\end{subequations}
with the spherical harmonics $Y_{\ell m}(\theta,\phi)$, where
$\ell$ and $m$ are integers satisfying $\ell\ge 0$ and $-\ell\le m\le \ell$.
Here, the first-order equations with different $(\ell, m)$ values
decouple, and each mode can be treated separately.
In what follows, we consider a single mode 
and write the radial functions as $H^{(1)}$ and $K^{(1)}$
for brevity.

As discussed in \cite{Chandrasekhar}, the radial equations for a fixed
$\ell$ have the same form for arbitrary $m$ 
due to the spherical symmetry of the background spacetime.
In the case of $m=0$, 
the first equality in Eq.~\eqref{function_nu1_mu1}
is found from the difference between the
$\theta\theta$ and $\phi\phi$ components of
the Einstein equations \eqref{Eq:vacuum-Einstein}. 
We consider the case $\ell\ge 2$ because the 
$\ell=0$ and $\ell=1$ modes correspond to coordinate transformations and
shift of the mass value. 
The radial equations for the first-order quantities are
\begin{subequations}
\begin{equation}
 r^2e^{2\nu^{(0)}}H^{(1)}_{,rr}
+ 2r\left(re^{2\nu^{(0)}}\right)_{,r} H^{(1)}_{,r}
- r^2\left(e^{2\nu^{(0)}}\right)_{,r}K^{(1)}_{,r} 
- \ell(\ell+1) H^{(1)} = 0,
\label{Rtt}
\end{equation}
\begin{multline}
 r^2e^{2\nu^{(0)}}\left(H^{(1)}_{,rr}-2K^{(1)}_{,rr}\right)
+ 2r\left( re^{2\nu^{(0)}}\right)_{,r} H^{(1)}_{,r} \\
- r\left[ r\left(e^{2\nu^{(0)}}\right)_{,r}+4e^{2\nu^{(0)}}\right] K^{(1)}_{,r}
+ \ell(\ell+1) H^{(1)} = 0,
\label{Rrr}
\end{multline}
\begin{eqnarray}
e^{2\nu^{(0)}}\left(H^{(1)}_{,r} - K^{(1)}_{,r}\right)
 + \left(e^{2\nu^{(0)}}\right)_{,r}H^{(1)} = 0,   
\label{Rr theta}
\end{eqnarray}
\begin{multline}
 r^2e^{2\nu^{(0)}} K^{(1)}_{,rr}
- 2re^{2\nu^{(0)}}H^{(1)}_{,r}
+ r\left[ r\left(e^{2\nu^{(0)}}\right)_{,r}+4e^{2\nu^{(0)}}\right] K^{(1)}_{,r} \\
- 2\left( re^{2\nu^{(0)}} \right)_{,r} H^{(1)} 
-  (\ell^2+ \ell  -2)K^{(1)} = 0,
\label{R phi phi}
\end{multline}
\end{subequations}
where ${,r}$ denotes the derivative with respect to $r$. 
For $m=0$, these equations are derived from $tt, rr, r\theta$ components and
the sum of the $\theta\theta, \phi\phi$ components of the Einstein equations,
respectively. 

It is convenient to introduce a normalized
radial coordinate as
\begin{equation}
x=\frac{r}{2M}.
\end{equation}
From Eqs.~\eqref{Rtt} and \eqref{Rrr},
the equation for $H^{(1)}$ is derived as
\begin{equation}
  x(x-1)H^{(1)}_{,xx}+(2x-1)H^{(1)}_{,x}
  -\left[\ell(\ell+1)+\frac{1}{x(x-1)}\right]H^{(1)} =0.
\end{equation}
The solution to this equation is\footnote{This solution was first
  presented in Eq.~(35) of Ref.~\cite{Regge:1957}, but a typo is included:
  this formula is not for $M$ but for $H$.
}
\begin{equation}
H^{(1)}(x) = \alpha_\ell P_{\ell}^2(2x-1) + \beta_{\ell}Q_{\ell}^{2}(2x-1),
\label{analytic-solution}
\end{equation}
where $P_{\ell}^{\mu}(z)$ and $Q_{\ell}^{\mu}(z)$
denote the associated Legendre functions of the first and second kinds,
\begin{subequations}
\begin{eqnarray}
  P_{\ell}^{\mu}(z) &=&
  \frac{1}{\Gamma(1-\mu)}\left(\frac{1+z}{1-z}\right)^{\mu/2}
       {}_2F_1\left(-\ell,\ell+1;1-\mu;\frac{1-z}{2}\right),\\
  Q_{\ell}^{\mu}(z) &=&
  \frac{\sqrt{\pi}\Gamma(\ell+\mu+1)}{2^{\ell+1}\Gamma(\ell+3/2)}\frac{(1-z^2)^{\mu/2}}{z^{\ell+\mu+1}}
       {}_2F_1\left(\frac{\ell+\mu+1}{2},\frac{\ell+\mu+2}{2};\ell+\frac32;\frac{1}{z^2}\right).
\end{eqnarray}
\end{subequations}
Here, ${}_2F_1(a,b;c;z)$ is the Gauss hypergeometric function.
Once $H^{(1)}$ is obtained, 
$K^{(1)}$ is calculated by the formula 
\begin{equation}
K^{(1)}(x) = \frac{1}{(\ell^2+\ell-2)} 
\left[
H^{(1)}_{,x}
+ \left(
  \frac{1}{x-1}+\frac{1}{x}+\ell^2+\ell-2
\right)H^{(1)}
\right],
\label{K1}
\end{equation}
which is derived by eliminating $H^{(1)}_{,rr}$, $K^{(1)}_{,rr}$ and $K^{(1)}_{,r}$ 
from Eqs.~\eqref{Rtt}--\eqref{R phi phi}.

The first term $\alpha_\ell P_{\ell}^2(2x-1)$
of the solution~\eqref{analytic-solution}
is regular on the horizon $x=1$ but is divergent
at infinity, $x\to \infty$, while the second term
$\beta_\ell Q_{\ell}^2(2x-1)$ 
decays at infinity but is divergent
on the horizon. 
In this paper, we consider distortion of
a Schwarzschild spacetime in the region outside
of the photon surface, $r\gtrsim 3M$.
When the spacetime is vacuum in this region,
we have to set $\alpha_\ell=0$ and 
adopt only the second term $\beta_\ell Q_{\ell}^2(2x-1)$
to make the perturbation regular.
Such a perturbation would be generated by, e.g., nonspherical distribution
of matter or naked singularities in the region $r<3M$. 
On the other hand, if we consider the distortion of a Schwarzschild
spacetime due to matter outside of $r=3M$,
the perturbation in the neighborhood of $r=3M$ is represented
by both terms of Eq.~\eqref{analytic-solution}
with nonvanishing $\alpha_\ell$ and $\beta_\ell$.

%
%
\section{Distortion of photon surface}

In this section, we examine whether a static photon surface
exists in a distorted Schwarzschild spacetime.

\subsection{Photon surface condition}

We rewrite the photon surface condition $\chi_{ab}\propto h_{ab}$
with perturbative quantities. 
Let a static photon surface in a distorted Schwarzschild spacetime be given as 
\begin{equation}
  r=f(\theta,\phi),
  \label{photon-surface-1}
\end{equation}
where
\begin{equation}
  f = f^{(0)} + \epsilon f^{(1)} +\cdots, \qquad \textrm{with} \qquad f^{(0)}=3M.
  \label{photon-surface-2}
\end{equation}
Here, $f^{(1)}$ is assumed to be dependent on the angular coordinates,
since the perturbation of the Schwarzschild spacetime has the possibility
to cause the distortion of the coordinate shape of the photon surface.

For this surface, we calculate the photon surface condition,
$\chi_{ab} \propto h_{ab}$.
We present the details of calculations in the Appendix
and just show the results here:
\begin{subequations}
\begin{eqnarray}
  f^{(1)}_{,\theta\phi}&=&\cot\theta f^{(1)}_{,\phi},
  \label{ps-eq1}
  \\
  f^{(1)}_{,\phi\phi}&=&\sin^2\theta f^{(1)}_{,\theta\theta} -\sin\theta\cos\theta f^{(1)}_{,\theta},
  \label{ps-eq2}
\end{eqnarray}
\begin{equation}
  \left.\left(\nu^{(1)}_{,r}-\psi^{(1)}_{,r}\right)\right|_{r=3M}
  =\frac{1}{3M^2}\left(f^{(1)}-f^{(1)}_{,\theta\theta}\right).
  \label{ps-eq3}
\end{equation}
\end{subequations}
The first two equations \eqref{ps-eq1} and \eqref{ps-eq2} are 
solved as
\begin{equation}
  f^{(1)} = \sin\theta \left(\alpha e^{i\phi} + \beta e^{-i\phi}\right)
  + \gamma\cos\theta + \delta,
  \label{shift-expand}
\end{equation}
where $\alpha$, $\beta$, $\gamma$ and $\delta$ are integral
constants.\footnote{
  The solution \eqref{shift-expand} 
  means that if the umbilical condition $\chi_{ab}\propto h_{ab}$ is restricted to
  the spatial part, allowed distortion is shift of the position toward three
  directions and uniform expansion. }
This is a linear combination of the
four spherical harmonics $Y_{\ell m}(\theta,\phi)$
with $(\ell,m) = (1,\pm 1)$, $(1,0)$, and $(0,0)$.
Then, the right-hand side of Eq.~\eqref{ps-eq3} has the $\ell = 0$ and $1$
modes, while the left-hand side has the $\ell\ge 2$ modes from the assumption.
Therefore, each side of Eq.~\eqref{ps-eq3} becomes zero,
and the right-hand side gives
\begin{equation}
f^{(1)} =0.
\end{equation}
Namely, in the Regge-Wheeler gauge, the coordinate
position of a distorted photon surface, if it exists, 
must remain at $r=3M$ to first order.
The left-hand side implies that
the metric perturbation must satisfy
\begin{equation}
  \left.\left(\nu^{(1)}_{,r}-\psi^{(1)}_{,r}\right)\right|_{r=3M} = 0
  \label{ps-condition-metric}
\end{equation}
in order for $r=3M$ to be a photon surface.

\subsection{Vacuum case}

Equation~\eqref{ps-condition-metric} indicates that
a photon surface can exist only when 
the metric perturbation satisfies a special property.
Here, we examine whether this property is satisfied
if the spacetime is vacuum outside of $r=3M$.

Rewriting 
with $H^{(1)}$ and $K^{(1)}$ and using Eq.~\eqref{Rr theta},
Eq.~\eqref{ps-condition-metric} is rewritten as
\begin{equation}
\left.\frac{H^{(1)}_{,\tilde{x}}}{H^{(1)}}\right|_{\tilde{x}=2} = -\frac13,
  \label{ps-condition-metric-2}
\end{equation}
where we introduced $\tilde{x} = 2x-1$ with $x=r/2M$.
The assumption that the region outside of $r=3M$ is vacuum corresponds
to $\alpha_{\ell} = 0$ in the solution of $H^{(1)}$ in Eq.~\eqref{analytic-solution}.
In this situation, the left-hand side of Eq.~\eqref{ps-condition-metric-2}
is calculated as
\begin{subequations}
\begin{equation}
  \left.\frac{\frac{d}{d\tilde{x}}Q^2_{\ell}(\tilde{x})}{Q^2_{\ell}(\tilde{x})}\right|_{\tilde{x}=2}
  =-\frac{3\ell+1}{6}
  -\frac14\left[\frac{\frac{d}{dz}{}_2F_1(a,b;c;z)}{{}_2F_1(a,b;c;z)}\right]_{z=1/4}
  \label{QL2DdbQL2}
\end{equation}
with 
\begin{equation}
a=\frac{\ell+3}{2}, \qquad b=\frac{\ell}{2}+2, \qquad c=\ell+\frac32.
\end{equation}
\end{subequations}
The explicit values of the formula~\eqref{QL2DdbQL2}
are shown in Table~\ref{Table:1} for $\ell = 2$,..., $5$.
Because both ${}_2F_1(a,b,c;z)$ and $\frac{d}{dz}{}_2F_1(a,b;c;z)$
are positive at $z=1/4$, 
the values of the formula~\eqref{QL2DdbQL2}
are smaller than $-1/3$ for arbitrary $\ell\ge 2$.
Therefore, the condition~\eqref{ps-condition-metric-2}
cannot be satisfied if the outside region is vacuum.

\begin{table}[tb]
  \caption{ The values of Eq.~\eqref{QL2DdbQL2}, $\left.\frac{d}{d\tilde{x}}Q^2_{\ell}(\tilde{x})/Q^2_{\ell}(\tilde{x})\right|_{\tilde{x}=2}$, in the vacuum case and the values of  
$\alpha_\ell/\beta_\ell$ to satisfy the condition~\eqref{ps-condition-metric-2}
  in the presence of matter  for $\ell=2,..., 5$.}
\begin{ruledtabular}
\begin{tabular}{c|rlrl}
  ~~$\ell$~~ & $\left.\frac{d}{d\tilde{x}}Q^2_{\ell}(\tilde{x})/Q^2_{\ell}(\tilde{x})\right|_{\tilde{x}=2}$ & (Numerical Value) & $\alpha_\ell/\beta_\ell$& (Numerical Value)  \\
  \hline
2  & $\frac{108\cdot\mathrm{arctanh}(1/2)-64}{81\cdot\mathrm{arctanh}(1/2)-42}$& $(-1.875)$ & $\frac{26}{45}-\mathrm{arctanh}\left(\frac{1}{2}\right)$ &
 ($2.847\times 10^{-2}$) \\
3  & $\frac{1485\cdot\mathrm{arctanh}(1/2)-818}{810\cdot\mathrm{arctanh}(1/2)-444}$& $(-2.431)$ & $\frac{322}{585}-\mathrm{arctanh}\left(\frac{1}{2}\right)$ &
 ($1.121\times 10^{-3}$) \\
4  & $\frac{25920\cdot\mathrm{arctanh}(1/2)-14240}{10935\cdot\mathrm{arctanh}(1/2)-6006}$& $(-2.995)$ & $\frac{5414}{9855}-\mathrm{arctanh}\left(\frac{1}{2}\right)$ &
 ($5.966\times 10^{-5}$) \\ 
5  & $\frac{182385\cdot\mathrm{arctanh}(1/2)-100186}{62370\cdot\mathrm{arctanh}(1/2)-34260}$& $(-3.563)$ & $\frac{37202}{67725}-\mathrm{arctanh}\left(\frac{1}{2}\right)$ & ($3.564\times 10^{-6}$)
\end{tabular}
\end{ruledtabular}
\label{Table:1}
\end{table}

The result here indicates that once a Schwarzschild spacetime
is distorted due to matter/naked singularities
inside of $r=3M$, the photon surface vanishes.
Therefore, the perturbative uniqueness holds 
for a spacetime with a static photon surface in 
the case that the outside region is vacuum.

\subsection{Nonvacuum case}

Here, we discuss the case that the outside
spacetime is not vacuum
and the distortion of a photon surface is caused
by matter distribution outside of $r=3M$,
as well as that inside of $r=3M$.
In such a case, the perturbation in the vicinity of
$r=3M$ is given by Eq.~\eqref{analytic-solution} with nonvanishing
$\alpha_\ell$ and $\beta_\ell$.
The introduction of the extra parameter $\alpha_\ell$ enables one
to satisfy 
the condition~\eqref{ps-condition-metric-2}. Namely,
the condition~\eqref{ps-condition-metric-2} becomes
an equation for $\alpha_{\ell}/\beta_\ell$, and 
the explicit values of the solutions are listed
in Table~\ref{Table:1} for $\ell = 2, ..., 5$.

Perturbative distortion of a photon surface 
is possible, in principle, 
if we discard the vacuum assumption outside of $r=3M$.
However, the requirement for  fine-tuning of $\alpha_\ell/\beta_\ell$
indicates that one has to distribute
matter in a very special manner 
in order to realize a distorted
photon surface.

%
%
\section{Conclusion}
\label{Sec:conclusion}

In this paper, we have examined whether a distorted Schwarzschild
spacetime can possess a static photon surface.
The photon surface condition $\chi_{ab}\propto h_{ab}$
requires the metric perturbation to have a special property
described by Eq.~\eqref{ps-condition-metric}.
If the region outside of $r=3M$ is vacuum,
the regular solution of the metric perturbation
does not satisfy this condition for any $\ell \ge 2$.
Therefore, we have proved the 
{perturbative uniqueness theorem: There is no solution
branch of a spacetime possessing a distorted photon surface
that regularly connects to the Schwarzschild solution, if the
outside spacetime is vacuum and asymptotically flat.}

From this perturbative uniqueness, two possibilities are indicated 
on the existence of a distorted photon surface.
One is that a spacetime solution
with a distorted photon surface exists but it does not
regularly connect to the Schwarzschild solution 
in the solution space.
The other is that there is no solution 
of a spacetime with a distorted photon surface at all.
Among these two, the first possibility is difficult to imagine
and would be unlikely. For this reason, we propose the following
{conjecture: If an asymptotically flat, vacuum spacetime
possesses a static photon surface, the spacetime is the Schwarzschild spacetime.}
Up to now, the uniqueness theorem has been proved
for a spacetime with a (generalized) photon sphere, i.e., a photon surface
given by a contour surface
of the time lapse function, $N=\sqrt{-g_{tt}}$.
Our result indicates that
it would be possible to eliminate the assumption
of the constancy of the lapse function on a photon surface
in proving the uniqueness theorem.
This direction is worth challenging.

On the other hand, if we consider a perturbation generated
by matter outside of $r=3M$, the perturbative distortion of a photon surface
has turned out to be possible although fine-tuning between
the amplitudes of two independent perturbative solutions is required.
The fine-tuning indicates that
matter has to be distributed in a special way in order
to realize a distorted photon surface.
Here, one has to keep in mind that the existence of a
first-order perturbative solution does not necessarily guarantee the
presence of a fully nonlinear solution because the nonlinearity may
invalidate it. Therefore,  
whether a photon surface can exist in the nonlinear regime
is an interesting  next issue, and it is worth
trying to construct explicit solutions.
Also, clarifying whether such a distorted photon surface exists
in a realistic astrophysical context would be an interesting remaining problem.

%
%
\acknowledgments

H.Y. thanks Shin Anase for helpful discussions that motivated this work.
This work was in part supported by the Grant-in-Aid for
Scientific Research (A) (No. 26247042)
from Japan Society for the Promotion of Science (JSPS).
We thank the referee of this paper
  for suggesting to add this section. There, we followed
  the logic strongly suggested by the referee.

\appendix
%
%
\section{Calculation of photon surface condition}
\label{Appendix_A}

In this appendix, we present calculations to derive
the perturbative photon surface condition, Eqs.~\eqref{ps-eq1}--\eqref{ps-eq3}.
Suppose a photon surface is given by $r=f(\theta,\phi)$ in a 
distorted Schwarzschild spacetime with a metric~\eqref{metric_deformed}.
We would like to calculate the photon surface condition
$\chi_{ab} \propto h_{ab}$
for this surface.
It is convenient to introduce coordinates $({\tilde{r}},\tilde{\theta},\tilde{\phi})$ such that
the photon surface is given by ${\tilde{r}}=0$ and
the radial coordinate basis $\partial_{\tilde{r}}$ is orthogonal
to ${\tilde{r}}=\mathrm{const.}$ slices. If such coordinates are introduced, 
the induced metric $h_{ab}$ on the ${\tilde{r}}=\mathrm{const.}$ slice
coincides with $g_{ab}$ 
except for the ${\tilde{r}}{\tilde{r}}$ component, and 
the extrinsic curvature is calculated by
\begin{equation}
\chi_{ab} = \frac{1}{2\sqrt{g_{{\tilde{r}}{\tilde{r}}}}}\partial_{\tilde{r}} h_{ab}.
\end{equation}
Then, the photon surface condition is equivalent to $\partial_{\tilde{r}} h_{ab}
\propto h_{ab}$.

For this reason, we consider the following coordinate transformation,
\begin{subequations}
\begin{eqnarray}
  r &=& {\tilde{r}}+f(\tilde{\theta},\tilde{\phi}),\\
  \theta &=& \tilde{\theta} - p({\tilde{r}},\tilde{\theta},\tilde{\phi}),\\
  \phi &=& \tilde{\phi} - q({\tilde{r}},\tilde{\theta},\tilde{\phi}),
\end{eqnarray}
\end{subequations}
and require $p=q=0$ on the photon surface so that $\theta=\tilde{\theta}$
and $\phi=\tilde{\phi}$ on ${\tilde{r}}=0$.
In order to make the metric components $g_{{\tilde{r}}\tilde{\theta}}$ and $g_{{\tilde{r}}\tilde{\phi}}$ vanish,
we further require
\begin{subequations}
\begin{eqnarray}
  f_{,\tilde{\theta}}e^{2\mu}
  &=&r^2e^{2\psi}\left[(1-p_{,\tilde{\theta}})p_{,{\tilde{r}}}-\sin^2\theta q_{,\tilde{\theta}}q_{,{\tilde{r}}}\right],
  \\
  f_{,\tilde{\phi}}e^{2\mu}
  &=&r^2e^{2\psi}\left[\sin^2\theta (1-q_{,\tilde{\phi}})q_{,{\tilde{r}}}-p_{,\tilde{\phi}}p_{,{\tilde{r}}}\right].
\end{eqnarray}
\end{subequations}
In particular, this relation is reduced to
\begin{subequations}
\begin{eqnarray}
  p_{,{\tilde{r}}}&=&\frac{f_{,\tilde{\theta}}}{f^2}e^{2(\mu-\psi)},\\
  q_{,{\tilde{r}}}&=&\frac{f_{,\tilde{\phi}}}{f^2}e^{2(\mu-\psi)}\sin^{-2}\theta,
\end{eqnarray}
\end{subequations}
on the photon surface ${\tilde{r}}=0$. Under these conditions on $p$ and $q$,
the nonvanishing metric components become
\begin{subequations}
\begin{eqnarray}
  g_{tt}&=&-e^{2\nu},\\
  g_{{\tilde{r}}{\tilde{r}}}&=&e^{2\mu}+r^2e^{2\psi}(p_{,{\tilde{r}}}^2+\sin^2\theta p_{,{\tilde{r}}}^2),\\
  g_{\tilde{\theta}\tilde{\theta}} &=&e^{2\mu}f_{,\tilde{\theta}}^2
  +r^2e^{2\psi}\left[(1-p_{,\tilde{\theta}})^2
  +\sin^2\theta q_{,\tilde{\theta}}^2\right],\\
  g_{\tilde{\phi}\tilde{\phi}} &=&e^{2\mu}f_{,\tilde{\phi}}^2
  +r^2e^{2\psi}\left[p_{,\tilde{\phi}}^2
  +\sin^2\theta (1-q_{,\tilde{\phi}})^2\right],\\
  g_{\tilde{\theta}\tilde{\phi}}&=&e^{2\mu}f_{,\tilde{\theta}}f_{,\tilde{\phi}}
  -r^2e^{2\psi}\left[(1-p_{,\tilde{\theta}})p_{,\tilde{\phi}}
  +\sin^2\theta (1-q_{,\tilde{\phi}})q_{,\tilde{\theta}}\right].
\end{eqnarray}
\end{subequations}

In what follows, we adopt the perturbative approximation and keep the
quantities up to the first order in $\epsilon$. For example,
since the derivatives of the metric components and the function $f$
with respect to the angular coordinates, $\tilde{\theta}$ and $\tilde{\phi}$, are
first order, their products (like $f_{,\tilde{\theta}}^2$ or $f_{,\tilde{\theta}}\mu_{,\tilde{\theta}}$)
are ignored.
The nonvanishing induced metric components on the
photon surface ${\tilde{r}}=0$ are
\begin{subequations}
\begin{eqnarray}
  h_{tt} &=& -e^{2\nu},\\
  h_{\tilde{\theta}\tilde{\theta}} &=& f^2e^{2\psi},\\
  h_{\tilde{\phi}\tilde{\phi}} &=& f^2e^{2\psi}\sin^2\theta.
\end{eqnarray}
\end{subequations}
On the other hand, the nonvanishing  components of $\partial_{\tilde{r}}h_{ab}$
on the ${\tilde{r}}=0$ surface are
\begin{subequations}
\begin{eqnarray}
  \partial_{\tilde{r}}h_{tt} &=& -2e^{2\nu}\nu_{,{\tilde{r}}},\\
  \partial_{\tilde{r}}h_{\tilde{\theta}\tilde{\theta}} &=&
  2f^2\left[e^{2\psi}\left(\frac{1}{f}+\psi_{,{\tilde{r}}}\right)-e^{2\mu}\frac{f_{,\tilde{\theta}\tilde{\theta}}}{f^2}\right],\\
  \partial_{\tilde{r}}h_{\tilde{\phi}\tilde{\phi}} &=&
  2f^2\left[e^{2\psi}\left(\frac{1}{f}+\psi_{,{\tilde{r}}}\right)\sin^2\theta
  -\frac{e^{2\mu}}{f^2}\left(f_{,\tilde{\phi}\tilde{\phi}}+\sin\theta\cos\theta f_{,\tilde{\theta}}\right)\right],\\
  \partial_{\tilde{r}}h_{\tilde{\theta}\tilde{\phi}} &=&
 -2e^{2\mu}\left(f_{,\tilde{\theta}\tilde{\phi}}-\cot\theta f_{,\tilde{\phi}}\right).
\end{eqnarray}
\end{subequations}
Since $h_{\tilde{\theta}\tilde{\phi}}=0$, the condition
$\partial_{\tilde{r}}h_{\tilde{\theta}\tilde{\phi}}\propto h_{\tilde{\theta}\tilde{\phi}}$
implies
$f_{,\tilde{\theta}\tilde{\phi}}=\cot\theta f_{,\tilde{\phi}}$,
which is identical to Eq.~\eqref{ps-eq1}.
The other conditions ${\partial_{\tilde{r}}h_{tt}}/{h_{tt}} = {\partial_{\tilde{r}}h_{\tilde{\theta}\tilde{\theta}}}/{h_{\tilde{\theta}\tilde{\theta}}} ={\partial_{\tilde{r}}h_{\tilde{\phi}\tilde{\phi}}}/{h_{\tilde{\phi}\tilde{\phi}}}$  
are written as
\begin{equation}
-f^2e^{2(\psi-\mu)}\left(\nu_{,{\tilde{r}}}-\psi_{,{\tilde{r}}}-\frac{1}{f}\right) = f_{,\tilde{\theta}\tilde{\theta}} = \frac{f_{,\tilde{\phi}\tilde{\phi}}}{\sin^2\theta}+\cot\theta f_{,\tilde{\theta}}.
\end{equation}
The second equality is equivalent to Eq.~\eqref{ps-eq2}.
Separating the first equality to the zeroth-order and first-order
equations taking account of Eq.~\eqref{photon-surface-2},
we find that the zeroth-order equation is satisfied
by $f^{(0)}=3M$, and the first-order equation is reduced
to Eq.~\eqref{ps-eq3}.

Although calculations become more tedious,
the perturbative photon surface condition
can be derived also in the original $(t,r,\theta,\phi)$ coordinates.
We have checked that
the same equations are obtained in this way.



\end{document}